\documentclass[aps,prc,showpacs,twocolumn]{revtex4}
\usepackage{graphicx}
\begin{document}

\def\pt{$p_T$ }
\def\ph{$\phi$ }
\def\om{$\Omega^-$ }
\def\rec{recombination }
\def\op{$\Omega/\phi$ }
\def\dis{distribution }

\title{Production of $\phi$ and $\Omega^-$ at RHIC in the Recombination Model}
\author{Rudolph C. Hwa$^1$ and  C.\ B.\ Yang$^{1,2}$}
\affiliation{$^1$Institute of Theoretical Science and Department
of Physics, University of Oregon, Eugene, OR 97403-5203, USA}
\affiliation{ $^2$Institute of Particle Physics, Hua-Zhong Normal
University, Wuhan 430079, P.\ R.\ China}
\date{}
\begin{abstract}
  The recombination model is applied to the production of \ph and \om at all \pt in
  central Au+Au collisions. Since no light quarks are involved in the hadronization,
  those hidden-strange particles present a clean slate for the study of the role of
  strange quarks in large-\pt physics. We find that shower $s$ quarks have negligible
  effect for $p_T<6$ GeV/c, in which range the thermal $s$ quarks make the dominant
  contributions to the formation of \ph and $\Omega^-$. We show that the same
  effective temperature of the $s$ quarks is responsible for the shape of the spectra of
  both $\phi$ and $\Omega^-$. We predict that the ratio of
  \om to \ph will show a peak at $p_T\approx 6$ GeV/c due to the effect of the hard
  partons. We also give reasons on the basis of the \pt dependence that \ph cannot
  be formed by means of $K^+K^-$ coalescence.

\end{abstract}
\maketitle

 \section{Introduction}
 The production of strange particles has always been a subject of great interest in
 heavy-ion collisions because of their relevance to possible signatures of deconfinement
 and flavor equilibration \cite{1,2}. Strangeness enhancement that has been observed at
 various colliding energies is a phenomenon associated with soft particles in the bulk
 matter \cite{3,4}. At high transverse momentum ($p_T$), on the other hand, the production
 of jets does not favor strange particles, whose fragmentation functions are suppressed
 compared to those for non-strange particles. At intermediate \pt range between the two
 extremes the \pt distribution depends sensitively on both the strangeness content and
 the production mechanism. It has been shown that in that \pt range the spectra of
 $\pi, K$, and $p$ can be well described by parton recombination \cite{5}. In this
 paper we study the production  of \ph and $\Omega^-$, both of which consist of only
 strange quarks. Without the participation of the non-strange quarks, they present a
 clean platform for the examination of the transverse momentum spectra of the $s$ and
 $\bar s$ quarks. Our aim is to learn about the transition from the enhanced thermal
 quarks to the suppressed shower quarks in the strange sector.

 Since the hidden-strange particles ($s\bar s, sss$) are expected to have small hadronic
 cross sections due to the OZI rule \cite{6, 7, 8}, \ph and \om are less likely to be
 affected by final-state interaction with co-movers in heavy-ion collisions, compared to
 kaons and hyperons. Their spectra should therefore reveal more directly their formation
 mechanism. We shall apply the recombination model, as in \cite{5}, and predict the shapes
 of their spectra beyond the intermediate \pt range where data do not yet exist. We can
 also calculate the $\Omega/\phi$ ratio and show  how different it is from the $p/\pi$
 ratio that has been the definitive signature of parton \rec  \cite{9,10,11}. If the
 prediction of the peaking of the $\Omega/\phi$ ratio at $p_T\approx 6$ GeV/c is verified
 by experiments, it will be another piece of evidence in support of \rec at high $p_T$.

 It should be emphasized that the focus of our work is on the hadronization of partons and
 not on the hydrodynamical evolution of the initial dense system. Thus we make no assertions
 on the nature of the thermal partons, but take their spectra from low-\pt data
 phenomenologically in the framework of the \rec model. The shower parton distributions
 are known from previous study \cite{12}, so our calculation of the hidden-strange hadronic
 spectra involves the \rec of various combinations of thermal and shower strange quarks.
 The enhanced thermal $s$ and the suppressed shower $s$ lead to an interesting departure
 from the more familiar scenario in the non-strange sector.

 \section{Formulation of the Problem}

 We shall assume that all hadrons produced in heavy-ion collisions are formed by \rec of
 quarks and/or antiquarks, the original formulation of which is given in \cite{13,14}.
 Later improvements are described in \cite{15} for hadronic collisions, in \cite{16} for
 $pA$ collisions and in \cite{5} for $AA$ collisions. For any colliding system the invariant
 inclusive distribution of a produced meson with momentum $p$ in a 1D description of the
 \rec process is
 \begin{eqnarray}
 p^0{dN_M  \over  dp} = \int {dp_1 \over  p_1}{dp_2
\over p_2}F_{q\bar{q}'} (p_1, p_2) R_M(p_1, p_2, p) ,
\label{1}
\end{eqnarray}
 and for a produced baryon
\begin{eqnarray}
p^0{dN_B\over dp}=\int {dp_1\over p_1}{dp_2\over p_2}{dp_3\over p_3}\,
F_{qq'q''}(p_1,p_2,p_3)\,R_B(p_1,p_2,p_3,p) .
\label{2}
\end{eqnarray}
 The properties of the medium created by the collisions are imbedded in the joint quark
 distributions $F_{q\bar q'}$ and $F_{qq'q''}$. The \rec functions (RF) $R_M$ and $R_B$
 depend on the hadron structure of the particle produced, but not on the medium out of
 which quarks hadronize. In the valon model description of hadron structure \cite{14,17},
 the RFs are
 \begin{eqnarray}
 R_M(p_1, p_2, p) =g_{M}\,y_1y_2\,  G_M (y_1 y_2 ) ,  \label{3}\\
 R_B(p_1,p_2,p_3,p)=g_{B}\,y_1y_2y_3\,G_B(y_1,y_2,y_3) , \label{4}
\end{eqnarray}
 where $y_i=p_i/p$, and $g_M$ and $g_B$ are statistical factors. $G_M$ abd $G_B$ are the
 non-invariant probability densities of finding the valons with momentum fractions $y_i$
 in a meson and a baryon, respectively.

    Equations (\ref{1}) and (\ref{2}) apply for the produced hadron having any momentum
    $\vec p$. We consider $\vec p$ only in the transverse plane, and write $p_T$ as $p$
    so that $dN/p_Tdp_T$ becomes $(pp^0)^{-1}$ times the right-hand sides of Eqs.\ (\ref{1})
    and (\ref{2}). For \ph and \om production, $F_{q\bar q'}$ is $F_{s\bar s}$, and
    $F_{qq'q''}$ is $F_{sss}$. The RFs are very narrow in momentum space, since both
    \ph and \om are loosely bound systems of the constituent $s$ quark. We shall
    approximate $G_\phi$ and $G_\Omega$ by $\delta$-functions:
\begin{eqnarray}
G_\phi(y_1,y_2)&=&\delta(y_1-1/2)\,\delta(y_2-1/2) ,   \label{5} \\
G_\Omega(y_1,y_2,y_3)&=&\delta(y_1-1/3)\,\delta(y_2-1/3)\,\delta(y_3-1/3) ,   \label{6}
\end{eqnarray}
and set $g_\phi=3/4$ and $g_\Omega=1/2$ for $J=1$ and $J=3/2$, respectively.
 Using these in Eqs.\ (\ref{3}) and (\ref{4}), and then in (\ref{1}) and (\ref{2}), we
 obtain the simple algebraic expressions
 \begin{eqnarray}
{dN_\phi\over pdp}&=&{3\over 4pp_0}\,F_{s\bar s}(p/2, p/2) ,   \label{7}\\
{dN_\Omega\over pdp}&=&{1\over 2pp_0}\,F_{sss}(p/3, p/3, p/3) ,  \label{8}
\end{eqnarray}
where $p_0=(m^2+p^2)^{1/2}$, $m$ being the mass of \ph or $\Omega^-$, as the case may be.

 In \cite{5} we have described how the joint parton distributions can receive contributions
 from the thermal ($\cal T$) and shower ($\cal S$) sources. In a schematic way they can be
 expressed as
 \begin{eqnarray}
F_{s\bar s}&=&{\cal TT+TS+SS},   \label{9}\\
F_{sss}&=&\kappa\,({\cal TTT+TTS+TSS+SSS}),   \label{10}
\end{eqnarray}
where showers from more than 1 jet are neglected. In Eq.\ (\ref{10}) the multiplicative factor
$\kappa$ is added to the $sss$ distribution to allow for the possible constraint arising from
the competition among various channels of hadronization that can limit the number of $s$
quarks available for the formation of $\Omega^-$.

 We parameterize the invariant $s$ quark \dis in the thermal source as
 \begin{eqnarray}
 {\cal T}(p_1) = p_1{dN_s^{\rm th}\over dp_1} = C_s\, p_1\exp (-p_1/T_s), \label{11}
\end{eqnarray}
where $C_s$ and $T_s$ are two parameters to be determined by fitting the low-\pt data; they
are expected to be different from those in the non-strange sector.  The distribution  of
shower $s$ quark in central Au+Au collisions is, as in \cite{5},
\begin{eqnarray}
{\cal S}(p_2) = \xi \sum_i \int^{\infty}_{k_0}dk\, k\,
f_i(k)\, S^s_i (p_2/k)  ,
\label{12}
\end{eqnarray}
 where $S_i^s$ is the shower parton distribution (SPD) for an $s$ quark in a shower initiated
 by a hard parton $i$, $f_i(k)$ is the transverse-momentum ($k$) \dis of hard parton $i$ at
 midrapidity in central Au+Au collisions, and $\xi$ is the average fraction of hard partons
 that can emerge from the dense medium to hadronize. As in \cite{5}, $f_i(k)$ is taken from
 Ref.\ \cite{18}, $k_0$ is set at 3 GeV/c, and $\xi$ is found to be 0.07.

 Equations (\ref{7})-(\ref{12}) completely specify the problem. There are 3 parameters to
 vary to fit the \pt spectra of \ph and $\Omega^-$: they are $C_s, T_s$ and $\kappa$. All
 aspects of the semi-hard shower partons have been fixed by previous studies \cite{5}. The
 three parameters are all related to the soft thermal partons, the properties of which, we
 have stated from the outset, are to be determined phenomenologically.

 \section{Results}

 We first vary $C_s$ and $T_s$ to fit the data on the \pt \dis of \ph in central Au+Au
 collisions at $\sqrt s=200$ GeV \cite{19}. For \pt up to 3 GeV/c, which is the extent
 to which data exist, the entire \dis can be accounted for by the \rec of the thermal
 partons only. The thermal-thermal component is shown by the dashed line in Fig.\ 1. The
 values of $C_s$ and $T_s$ used for the fit  are
 \begin{eqnarray}
C_s=8.64\ {\rm (GeV/c)^{-1}}, \qquad\qquad T_s=0.385\ {\rm GeV/c}.   \label{13}
\end{eqnarray}
The contributions from thermal-shower (dash-dot line) and
shower-shower (line with crosses) \rec do not become important
until $p_T\approx 7$ GeV/c. The solid line in Fig.\ 1 indicates
the sum of all contributions. The dominance of the thermal
component for  $p_T<6$ GeV/c is due to the fact that the
production of $s$ quarks in a shower is highly suppressed
\cite{12}. For that reason the role of the shower partons in the
formation of \ph in the intermediate \pt region is insignificant
compared to that for the production of pions \cite{5}.
Shower-shower recombination that can be related to fragmentation
can become important, but not until $p_T>8$ GeV/c.

\begin{figure}
\includegraphics[width=0.45\textwidth]{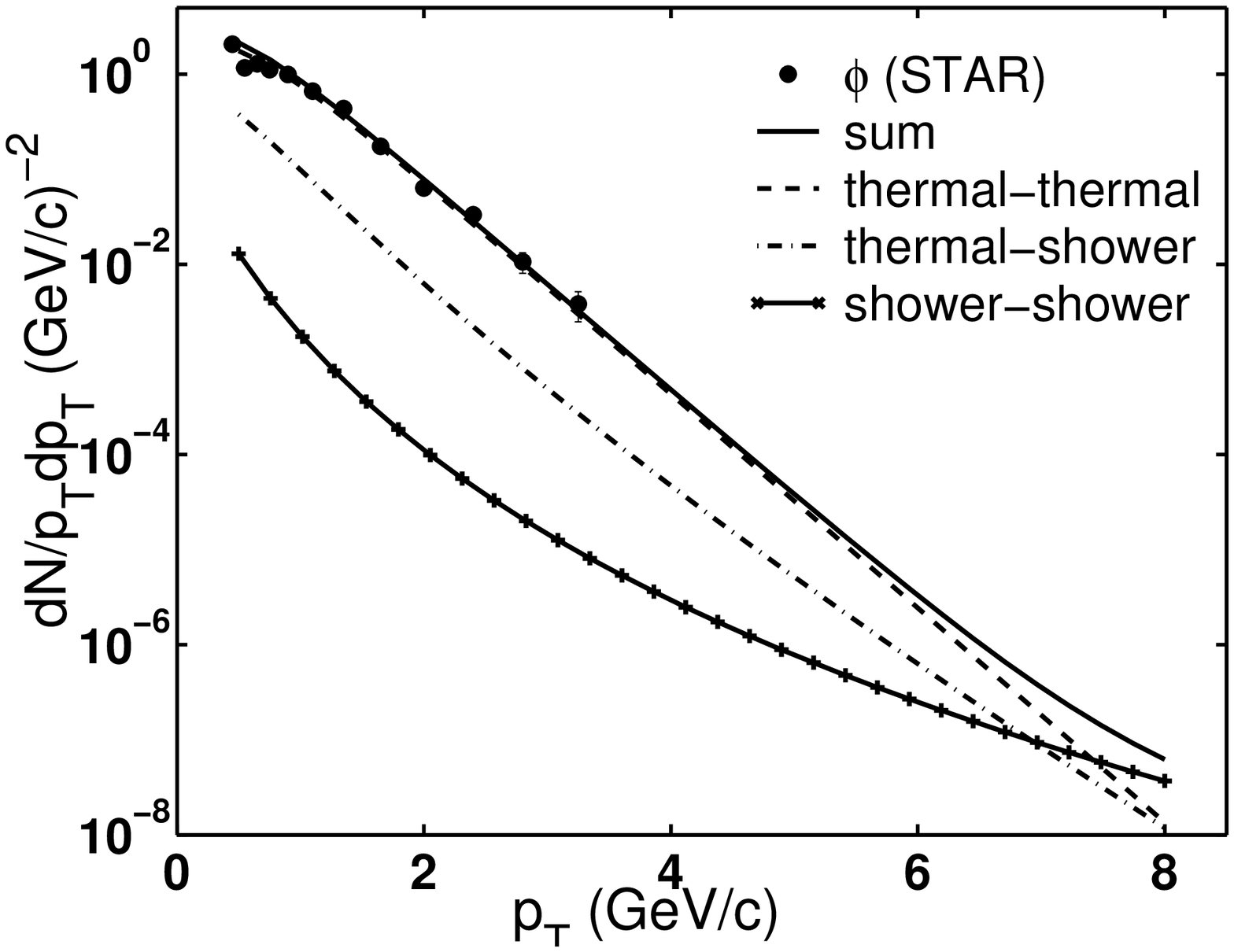}
\caption{ Transverse momentum distribution of \ph in central Au+Au
collisions. Data are from \cite{19}. The solid line is the sum of
the three contributions: $\cal TT$ (dashed line); $\cal TS$
(dash-dot line);  $\cal SS$ (line with crosses). }
\end{figure}

Next, for the production of \om there is only one parameter
$\kappa$ to vary. Again, the  \rec of thermal partons only
dominates throughout the whole \pt region shown in Fig.\ 2. All
other contributions that involve the participation of at least one
shower parton are increasingly negligible with increasing number
of shower $s$ quark. The \rec of $\cal SSS$ can be identified with
the fragmentation of hard partons to \om in the same sense that
the \rec of only the shower $uud$ in a jet gives the proton by
fragmentation \cite{5}, even though the fragmentation function for
\om does not exist. Evidently, the contribution to the \om
spectrum from fragmentation is negligible until \pt is extremely
large.

\begin{figure}
\includegraphics[width=0.45\textwidth]{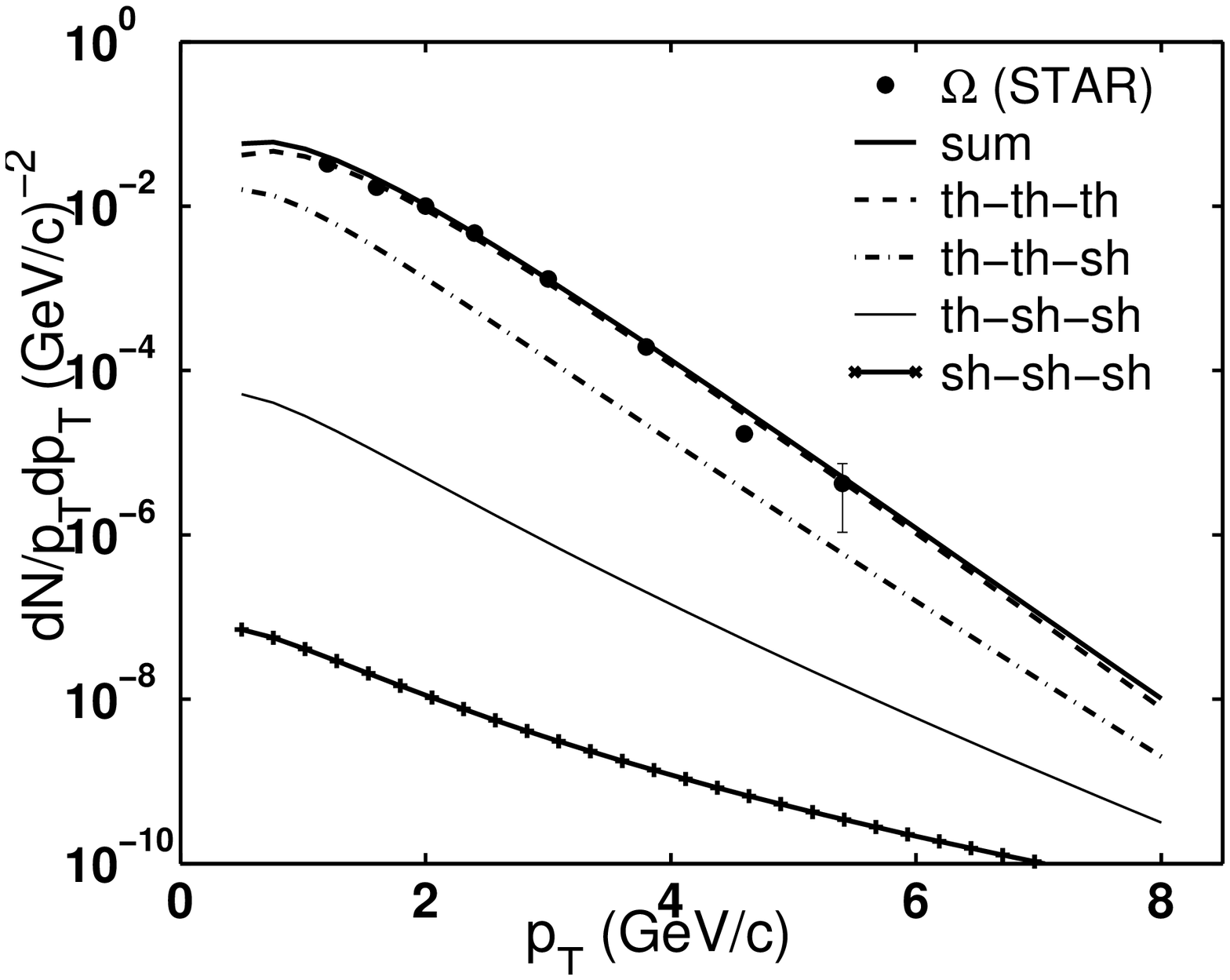}
\caption{ Transverse momentum distribution of \om in central Au+Au
collisions. Data are from \cite{20}. The heavy solid line is the
sum of the four contributions: $\cal TTT$ (dashed line); $\cal
TTS$ (dash-dot line); $\cal TSS$ (light solid line); $\cal SSS$
(line with crosses).}
\end{figure}

The value of $\kappa$ used in the fit of the data \cite{20} is
\begin{equation}
\kappa=0.037 .     \label{14}
\end{equation}
It affects only the overall normalization of the spectrum, not the
relative magnitudes  of the different components. Thus our
predictive power is in the shape of \om spectrum, not in the
normalization. The agreement with the data in Fig. 2 is clearly
excellent. Apart from the different powers of $p_1$ in Eq.\
(\ref{11}) that appear in (\ref{9}) and (\ref{10}), the shape is
controlled by $T_s$, which is common for both \ph and \om
production. The low level of \om production is a phenomenological
fact that is embodied in the smallness of $\kappa$ in Eq.\
(\ref{14}). The underlying physics is most likely the demand for
$s$ quarks in the formation of lower-mass particles, such as
kaons, hyperons and \ph meson, so that their availability for
forming the higher-mass \om is significantly reduced. To consider
all channels of hadronization simultaneously would require a study
similar to those carried out in \cite{21,22,23}, which is beyond
the scope of this work, where our emphasis is on the \pt
dependences of \ph and $\Omega^-$.

\begin{figure}
\includegraphics[width=0.45\textwidth]{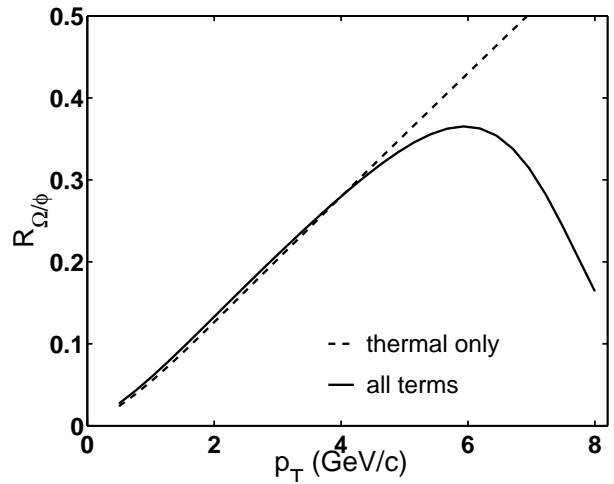}
\caption{ Transverse momentum dependence of the calculated
$\Omega/\phi$ ratio. The dashed line indicates the result from
thermal contributions only, while the solid line includes the
contributions from the shower $s$ quarks.}
\end{figure}

Having obtained \ph and \om spectra, let us present their ratio
$R_{\Omega/\phi}(p_T)$,  shown in Fig.\ 3. The dashed line is the
ratio of only the thermal contributions. It is linearly rising
because, on account of Eqs. (\ref{9}) and (\ref{10}), it is
proportional to  $p$ at large $p$ after the common exponential
terms of the two spectra are cancelled. The solid line shows the
effect when the thermal-shower \rec is taken into account.
Although the effect of shower partons on \om is small, the effect
on \ph above $p_T=5$ GeV/c is substantial enough to turn the
dashed line in Fig.\ 3 into a peak at around 6 GeV/c. Confirmation
of this peak by future experimental data would be a definitive
validation of the theoretical approach taken here and, in
particular, the importance of thermal-shower recombination.

\section {Discussion}

The strange sector differs from the non-strange sector in several
ways.  Firstly, the values of $C_s$ and $T_s$ given in Eq.\
(\ref{13}) are different from
\begin{equation}
C=23.2\ {\rm (GeV/c)^{-1}} ,  \qquad\qquad T=0.317\ {\rm GeV/c} ,  \label{15}
\end{equation}
for the non-strange quarks \cite{5}. Strangeness enhancement
refers to the excess of $s$ quarks in $AA$ collisions compared to
the scaled $pp$ collisions, but the total number of $s$ quarks is
still much less than that of light quarks in nuclear collisions.
That difference is reflected in $C_s$ being much less than $C$.
The inverse slope $T_s$ is, however, higher than $T$, since
hydrodynamical flow gives the more massive $s$ quark a larger mean
transverse momentum than it does the light quarks. The combined
effect of  larger $T_s$ and weaker shower $s$ quarks results in
the thermal partons becoming a dominant contributor to the
formation of \ph and \om over a much wider range of \pt than in
the case of non-strange quarks. For pion production the
thermal-shower \rec becomes important for $p_T>3$ GeV/c, whereas
for \ph it is not until $p_T>7$ GeV/c.

The above comment cannot be checked directly by experiments.
However, the  predicted ratio of \om to \ph can be tested. The
rising portion of that ratio in Fig.\ 3 is much broader than that
for the $p/\pi$ ratio \cite{24}, and is an indication of the
dominance of thermal-thermal recombination. The overall shape of
$R_{\Omega/\phi}$ is very different from that of $R_{p/\pi}$. The
latter peaks at $p_T\approx 3$ GeV/c with an experimental value
exceeding 1, whereas the former is predicted to peak at
$p_T\approx 6$ GeV/c with a value less than 0.4. Neither
$R_{p/\pi}$ nor $R_{\Omega/\phi}$ exhibits properties that can be
associated with fragmentation. The difference between them
reflects the differences between $u, d$ and $s$ quarks on the one
hand, and between non-strange and hidden-strange hadrons on the
other.

Within the framework of \rec it is possible for us to examine the
reality  of \ph formation through $K^+K^-$ coalescence, which is a
mechanism that has been advocated in certain models \cite{25,26}.
If $H_h(p_T)$ denotes the invariant inclusive \pt distribution,
$p^0dN_h/dp_T$, of hadron $h$ at $y=0$ in heavy-ion collisions,
then the coalescence process of $K^++K^-\rightarrow \phi$ implies
by use of Eq.\ (1) that
\begin{eqnarray}
H_\phi^{[KK]}(p_T) \propto H_K^2(p_T/2)   \label{16}
\end{eqnarray}
apart from a multiplicative constant associated with the RF. On
the other  hand, if \ph is produced by $s\bar s$ \rec as we have
done here, then the same procedure yields
\begin{eqnarray}
H_\phi^{[ss]}(p_T) \propto {\cal T}_s^2(p_T/2) ,   \label{17}
\end{eqnarray}
 where the dominance of the thermal parton \rec is used. Thus it is a
 matter of comparing $H_K(p)$ with ${\cal T}_s(p)$, which are the invariant distributions
 of the entities that recombine.
 Since a kaon is formed by $\bar sq$ recombination, where $q$ denotes either $u$ or $d$,
 whose thermal \dis ${\cal T}(p)$ is characterized by $C$ and $T$ shown in Eq.\ (\ref{15}),
 the exponential part of $H_K(p)$ must have an inverse slope $T'$ that is between $T$ and
 $T_s$, i.e., $~e^{-p/T'}$ with $2/T'=T^{-1}+T_s^{-1}$.  That is to be compared to
 the exponential part of ${\cal T}_s^2(p/2)={\cal T}_s(p)$, which is $e^{-p/T_s}$
 [cf. Eq.\ (\ref{11})]. In view of our good fit of the \ph data in Fig.\ 1, we conclude
 that an alternative fit using $H_K(p)$ characterized by $T'$ would fail. It therefore
 follows from the consideration of the \pt dependence alone that \ph cannot be formed by
 $K^+K^-$ coalescence. This conclusion is consistent with that of \cite{19} based on the
 centrality independence of the $\phi/K^-$ ratio.

 In summary, we have shown that both \ph and \om are formed mainly by the \rec of thermal
 $s$ quarks, have reproduced the shape of the \pt \dis of \om from a fit of that of $\phi$,
 and have made a prediction of the ratio of \om to \ph that peaks at $p_T\approx 6$ GeV/c.
 Thermal-shower \rec does not become important until \pt is at around that peak,
 and parton fragmentation does not dominate until $p_T$ is much higher. Finally,
 we have shown how the \pt dependence disfavors the formation of \ph by $K^+K^-$ coalescence.

\section*{Acknowledgment}
 We thank Nu Xu for his encouragement and communication in the pursuit of this problem.
 This work was supported, in part,  by the
U.\ S.\ Department of Energy under Grant No. DE-FG02-96ER40972
and by the Ministry of Education of China under Grant No. 03113.

\end{document}